# Cross-plane enhanced thermoelectricity and phonon suppression in graphene/MoS$_2$ van der Waals heterostructures

Hatef Sadeghi[*], Sara Sangtarash and Colin J. Lambert[*]

Quantum Technology Centre, Physics Department, Lancaster University, Lancaster LA1 4YB, UK

E-mail: [*]*h.sadeghi@lancastter.ac.uk; c.lambert@lancaster.ac.uk*

**Abstract.** The thermoelectric figures of merit of pristine two-dimensional materials are predicted to be significantly less than unity, making them uncompetitive as thermoelectric materials. Here we elucidate a new strategy that overcomes this limitation by creating multi-layer nanoribbons of two different materials and allowing thermal and electrical currents to flow perpendicular to their planes. To demonstrate this enhancement of thermoelectric efficiency *ZT*, we analyse the thermoelectric performance of monolayer molybdenum disulphide (MoS$_2$) sandwiched between two graphene monolayers and demonstrate that the cross-plane (CP) *ZT* is significantly enhanced compared with the pristine parent materials. For the parent monolayer of MoS$_2$, we find that *ZT* can be as high as approximately 0.3, whereas monolayer graphene has a negligibly small *ZT*. In contrast for the graphene/MoS$_2$/graphene heterostructure, we find that the CP *ZT* can be as large as 2.8. One contribution to this enhancement is a reduction of the thermal conductance of the van der Waals heterostructure compared with the parent materials, caused by a combination of boundary scattering at the MoS$_2$/graphene interface which suppresses the phonons transmission and the lower Debye frequency of monolayer MoS$_2$, which filters phonons from the monolayer graphene. A second contribution is an increase in the electrical conductance and Seebeck coefficient associated with molybdenum atoms at the edges of the nanoribbons.

## 1. Introduction

With a view to advancing materials performance beyond current silicon-CMOS [1], two dimensional materials [2–4] such as graphene [5], molybdenum disulfide (MoS$_2$) [6,7], phosphorene [8], silicene [9,10] and other transition metal dichalcogenides [11,12] have attracted considerable scientific interest, partly due to their potential for realising devices with sub-10-nanometre dimensions. Although graphene has attracted huge scientific interest, it behaves as a semimetal, which is useful as an electrode material [13], but not as a channel for electronic devices. On the other hand other two-dimensional materials, such as monolayer MoS$_2$ [6,7], possess band-gaps comparable with silicon and are attractive for current semiconducting technology. MoS$_2$ is a metal-dichalcogenide produced by exfoliation [14] of bulk MoS$_2$ or grown by CVD [15] and is resilient to oxidation. A number of recent studies have revealed how electronic properties change when bulk MoS$_2$ (with an indirect energy band-gap of about 1.2 $eV$) is exfoliated to create a monolayer structure [7,16] (with a 1.8 $eV$ direct band gap) and when such monolayers are stacked on graphene to form MoS$_2$/graphene [14,15,17–28] or graphene/MoS$_2$/graphene [29,30] heterostructures, thereby increasing their potential applications for field effect transistors and memory devices. In contrast their potential for thermoelectricity not yet fully characterized.

From the viewpoint of thermoelectricity, the in-plane thermal conductance of graphene [31] is higher than pristine MoS$_2$, which makes graphene useful for thermal management [32], but reduces its efficiency as a thermoelectric material. Since pristine monolayer MoS$_2$ has a lower thermal conductance, our aim below is to examine whether this material can form a basis for the design of high-efficiency thermoelectric materials. Compared with the in-plane thermal conductance of pristine monolayer MoS$_2$ [33,34], calculations show that cross-plane thermal transfer is less effective at physisorbed surfaces formed by MoS$_2$ and other metals due to weak interface interactions with MoS$_2$ [24]. Recently, molecular dynamics simulations showed that this interfacial thermal conductance is suppressed in a bulk MoS$_2$/graphite sandwich [35] and in hybrid monolayer structures [25] and therefore in what follows, we investigate whether or not this tendency leads to low-thermal-conductance and enhanced thermoelectric performance for cross-plane MoS$_2$ van der Waals heterostructures. In contrast with studies of phonon transport alone, our study below requires simultaneous evaluation of both electron and phonon transport, because heat is transported by both carriers [36,37]. Indeed as we show the thermal conductance could be dominated by electrons or phonons in the graphene/MoS$_2$/graphene van der Waals heterostructure depending on the device Fermi energy.

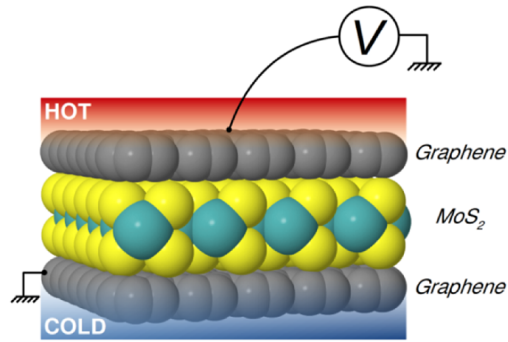

**Figure 1. The atomic structure of the graphene/MoS$_2$/graphene heterostructure.** The van der Waals distance between two layer is 3.4Å.

Electronic properties such as electrical conductance $G$, thermopower (Seebeck coefficient) $S$, and thermal conductance due to the electrons $\kappa_{el}$ can be obtained from the transmission probability $T_{el}(E)$ of electrons with energy $E$ passing from one electrode to another, whereas the thermal conductance due to the phonons $\kappa_{ph}$ is obtained from the transmission probability $T_{ph}(\omega)$ of the phonons with energy $\hbar\omega$ passing from one electrode to another [36] (see methods). In any thermoelectric device, the goal is to generate a voltage difference $\Delta V$ from a temperature difference $\Delta T$ between electrodes, from which the Seebeck coefficient is defined to be $S = -\Delta V/\Delta T$.

In addition, for the efficient conversion of heat into electricity, the dimensionless thermoelectric figure of merit $ZT = GS^2T/\kappa$ where $\kappa = \kappa_{el} + \kappa_{ph}$ should be maximised. Since $G$, $S$ and $\kappa_{el}$ are interrelated (see methods) and since the tuning of mechanical properties to minimise $\kappa_{ph}$ can also affect electrical properties, the goal of enhancing $ZT$ is a delicate optimization problem. The highest yet reported room temperature $ZT$ of 2.2 is found in $Bi_2Te_3/Sb_2Te_3$. However materials based on bismuth or tellurium are highly toxic and have a limited global supply. Hence these cannot provide a long term solution for future energy demands.

In this paper, we study the thermoelectric properties of graphene/MoS$_2$ van der Waals heterostructures (Gr/MoS$_2$) and for the first time, provide a detailed analysis of their thermal conductance due to both phonons and electrons. By taking advantage of electronic states associated with Mo atoms at the edges of such structures, we find that high thermopower and electrical conductance can be combined with a significant reduction in the heat transport leading to a high thermoelectric figure of merit $ZT$. Since we are interested in minimising the thermal conductance, we chose to study cross-plane (CP) transport (figure 1), since any device in which currents flow parallel to the planes [38] may suffer from parallel heat paths in a nearby substrate. Even in the absence of such parallel heat paths, we show that for the CP structure, the thermal conductance due to the phonons is significantly reduced. From a fabrication point of view, STM lithography yields graphene naonribbons of width 2.5 nm to 15 nm, and a length of ~120 nm [39], so we expect that these devices made from stacked layers of graphene and MoS$_2$ [27,30], could be similarly processed to form ribbons. The resulting prediction of a high $ZT$, combined with ease of fabrication and a cheap and plentiful global supply, suggest that these van der Waals heterostructures have high potential for future thermoelectric power generation.

## 2. Result and discussion

Figure 1 shows an atomic-scale schematic of the structure of the molybdenum disulphide/graphene (Gr/MoS$_2$) hetrostructure studied here. A single layer of MoS$_2$ is sandwiched between two single layers of graphene. As shown in figure S1 in the supporting information SI, periodic boundary conditions are applied in transverse (y) direction and the MoS$_2$ forms a ribbon of width $d$ in the transport (z) direction. In what follows, we present results for two ribbon widths $d = 1.9$nm and $d = 3.8$nm. Due to a small mismatch between the two pristine lattice constants, the Mo-Mo bonds in the sandwich of figure 1 shrink by approximately 0.01Å to match the graphene lattice. After geometry relaxation (see methods), the van der Waals distance between graphene and MoS$_2$ layers are 3.4Å, which is consistent with other calculations [40].

To demonstrate how the thermal properties of a sandwiched Gr/MoS$_2$ structure differs from those of pristine graphene and pristine MoS$_2$ monolayers, we first study the thermoelectric properties of the parent pristine monolayers and show that our findings are in good agreement with reported experimental results. Then we consider the van der Waals heterostructures and demonstrate that a much-more-efficient thermoelectric device is realised using these structures. We obtain the relaxed geometries of pristine monolayer graphene, pristine MoS$_2$ and the Gr/MoS$_2$ heterostructure of figure 1 using density functional theory (DFT). Their DFT mean field Hamiltonians

[41] were then combined with our transport code Gollum [42] to calculate the electronic transmission coefficient $T_{el}$, from which the thermopower S, the electronic thermal conductance $\kappa_{el}$ and electrical conductance G of each structure (see methods) were obtained. For phonon properties, DFT was also used to construct the dynamical matrix of the relaxed geometries, from which the phonon transmission coefficient $T_{ph}$ and phonon thermal conductance $\kappa_{ph}$ were obtained (see methods).

Figure 2 shows the electron and phonon transmission coefficients of the pristine structures. In agreement with other studies [13], graphene shows two open electronic conduction channels around the Fermi energy (figure 2a), whereas the MoS$_2$ monolayer possesses an energy gap [7] (figure 2b). We recently demonstrated that an asymmetric (with respect to the Fermi energy $E_F$) step-function-like transmission could produce a high thermopower, due to a high asymmetry of the transmission coefficient close to the Fermi energy [43]. Therefore, one would expect a high thermopower for the MoS$_2$, whereas the thermopower of graphene should be small, because the transmission coefficient is symmetric around Dirac point ($E_F$ =0). Indeed a high thermopower can be induced in CVD MoS$_2$ at room temperature by varying the back-gate voltage [44] and a low thermopower of ~50μV/K was observed in single-layer graphene at room temperature [32], in good agreement with our calculation here.

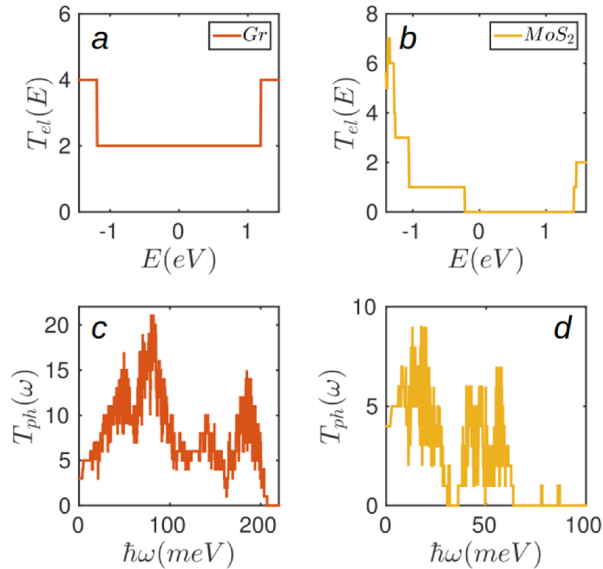

**Figure 2. Electron and phonon transport in pristine graphene and MoS$_2$.** (a,b) the electron transmission coefficient $T_{el}(E)$ and (c,d) the phonon transmission coefficient $T_{ph}(\hbar\omega)$.

Figures 2c,d show the phonon transmission coefficients $T_{ph}$ obtained from the dynamical matrix of the same DFT-relaxed geometries (see methods). We find that monolayer graphene possesses a high Debye temperature of $T_D$ ~ *2400K (ie 207 meV)*, in agreement with the literature value of $T_D$ = 2300K [45] (*ie $\hbar\omega_D = k_B T_D$ = 198.4 meV*). In contrast, we predict a much lower Debye frequency $\omega_D$ for a MoS$_2$ monolayer (about one quarter of the graphene Debye frequency). As we shall demonstrate below, in a sandwich structure, this means that the MoS$_2$ will filter the high frequency modes of the graphene at high temperatures. Figures 3a,b show the corresponding results for the thermal conductances of monolayer graphene and pristine MoS$_2$, while figures 3c,d show their thermopowers (ie Seebeck coefficients S). In agreement with experimental values [31] the thermal conductance of graphene is very high, which combines with the low graphene thermopower (figure 3c) to produce a low thermoelectric figure of merit ZT (figure 3d). However, in pristine MoS$_2$, the thermal conductance is mainly due to phonons, because the electronic contribution to thermal conductance at DFT-predicted Fermi energy is low (figure 3a,b). The thermal conductance in pristine monolayer MoS$_2$ is lower than graphene with similar size (figure 3a). Furthermore due to the asymmetric $T_{el}$ of monolayer MoS$_2$ (figure 2b), where the valance band is close to the DFT predicted Fermi energy, the thermopower is high over a wide range of Fermi energies around $E_F=0$ (figure 3c).

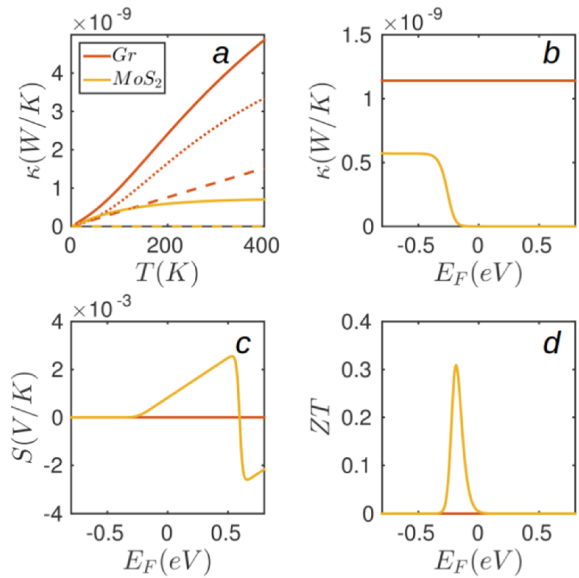

**Figure 3. Thermoelectric properties of pristine graphene and pristine MoS$_2$.** (a) The total thermal conductance $\kappa$ (solid line) and the separate electron $\kappa_{el}$ (dashed line) and phonon $\kappa_{ph}$ (dotted line) contributions. The Fermi energy dependence of the room temperature (b) electronic thermal conductance, (c) the thermopower S and (d) total thermoelectric figure of merit ZT in pristine Graphene and pristine MoS$_2$ monolayer. In (a) $\kappa_{el}$ is obtained at the DFT-predicted Fermi energy ($E_F = 0$).

High thermopower combined with fairly low thermal conductance in pristine monolayer MoS$_2$ lead to a ZT as high as 0.3 around the Fermi energy at room temperature, which is comparable with the best thermoelectric materials operating in the room temperature [46]. In what follows, our aim is to show that by sandwiching the MoS$_2$ between two graphene electrodes, much higher values of ZT are accessible.

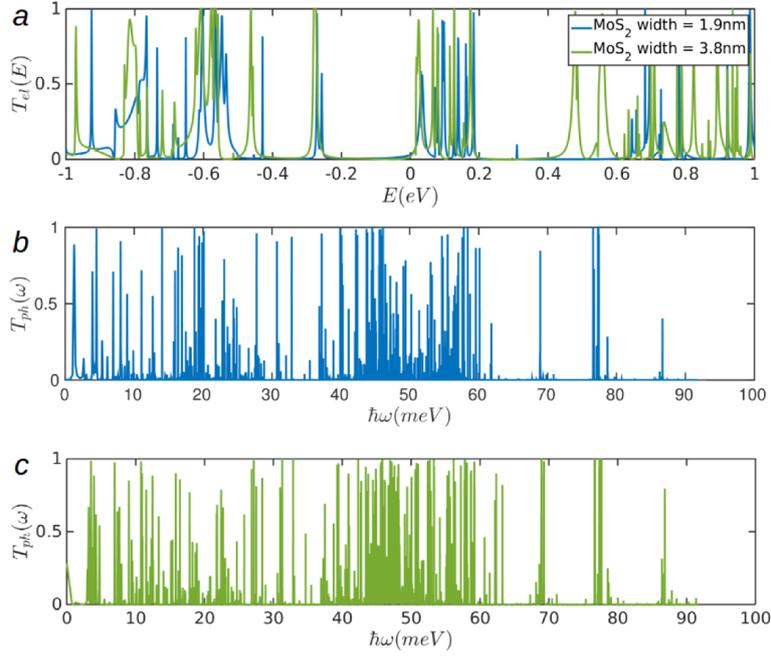

**Figure 4. Electron and phonon transport in the graphene/MoS$_2$/graphene heterostructure shown in figure 1**, formed from MoS$_2$ ribbons of width d=1.9nm (blue lines) and d = 3.8nm (green lines). (a) the electron transmission coefficient $T_{el}(E)$. (b) and (c) the phonon transmission coefficient $T_{ph}(\omega)$ for widths d=1.9nm (blue lines) and d = 3.8nm (green lines) respectively.

To study the thermoelectric properties of the graphene/MoS$_2$/graphene van der Waals heterostructure of figure 1, we first calculated the electron and phonon transmission coefficients of Gr/MoS$_2$/Gr, shown in figure 4. In an energy range of ~0.2eV above the DFT-predicted Fermi energy $E_F=0$, the electron transmission coefficient possesses a collection of sharp resonances (figure 4a). As shown by local density of states plots in figure S6, these resonances are due to states located on the edges of the ribbon and consequently their energies change only slightly as the ribbon width is doubled from 1.9nm to 3.8nm. As shown in the figure S5 of the SI, when such a delta-function-like transmission coefficients are located asymmetrically relative to the Fermi energy very high values of the electronic figure of merit $ZT_{el}= GS^2T/\kappa_e$ [37] can be expected.

As shown in figure 4b, the phonon transmission coefficient is strongly suppressed (figure S9) in Gr/MoS$_2$/Gr heterostructure compared with pristine graphene (figure 2c) and MoS$_2$ (figure 2d), because modes higher than the MoS$_2$ Debye frequency are filtered ($T_{ph} \sim 0$ for $\hbar\omega > \sim 90meV$). Furthermore, despite the presence of a high number of open phonon conduction channels in both graphene and MoS$_2$ monolayers, for phonon energies lower than ~90meV, the phonon transmission coefficient of the van der Waals heterostructure is strongly suppressed due to a high phonon mismatch between the graphene and MoS$_2$ monolayers interfaces in the cross-plane configuration.

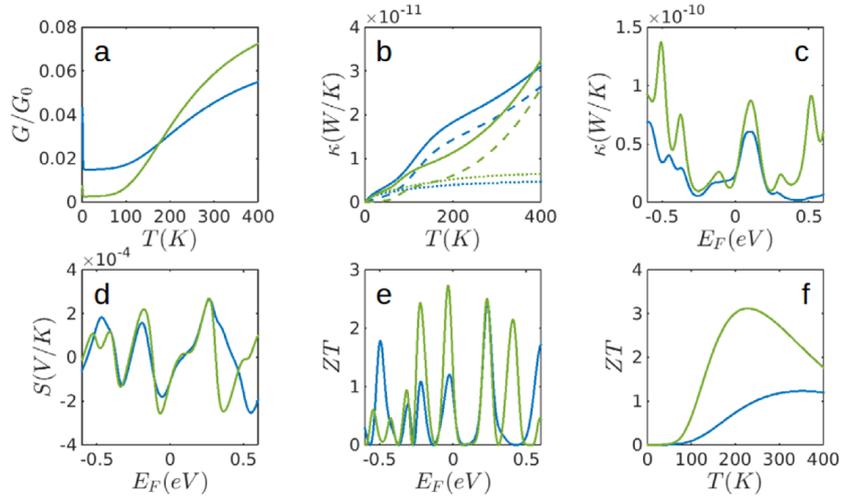

**Figure 5. Thermoelectric properties of the graphene/MoS$_2$/graphene heterostructure of figure 1.** Results are shown for MoS$_2$ ribbons of width d=1.9nm (blue lines) and d = 3.8nm (green lines) (a) The electrical conductance $G$ in units of the conductance quantum $G_0$. (b) The total thermal conductance $\kappa$ (solid lines) and the separate electron $\kappa_{el}$ (dashed lines) and phonon $\kappa_{ph}$ (dotted lines) contributions. The Fermi energy dependence of the room temperature (c) total thermal conductance, (d) the thermopower S and (e) total thermoelectric figure of merit $ZT$. (f) The figure of merit obtained at a Fermi energy of -0.03eV relative to the DFT-predicted Fermi energy. In (a) and (b) $G$ and $\kappa$ respectively are obtained at a Fermi energy of -0.03eV relative to the DFT-predicted Fermi energy.

From the graphene/MoS$_2$/graphene-heterostructure electronic and phonon transmission coefficients of figure 4, we calculated the electrical conductance, the thermal conductance due to both electrons and phonons, the thermopower and the figure of merit $ZT$. Figure 5 shows the resulting thermoelectric properties of Gr/MoS$_2$/Gr. Surprisingly the thermal conductance is mostly due to the electrons at the DFT-predicted Fermi energy as a consequence of a high electronic transmission coefficient (figures 4a and 5b), which increases the relative contribution of electrons to the total thermal conductance. This significant optimization of the electronic properties is decoupled from the phonon engineering of the Gr/MoS$_2$/Gr CP heterostructure, because the asymmetric electron transmission is partly due to molybdenum atoms at the edge of the Gr/MoS$_2$/Gr strip (see figure S6d,e of the SI), which creates sharp transmission resonances close to the DFT-predicted Fermi energy ($E_F$=0). Due to these asymmetric transmission features, a Gr/MoS$_2$/Gr Seebeck coefficient of approximately -200 $\mu V/K$ is obtained (figure 5d). This high thermopower combines with the low thermal conductance to yield a high room-temperature $ZT$ of approximately 1.2 (figure 5e) at room temperature in the vicinity of the DFT Fermi energy for the narrower ribbon and a higher value of approximately 1.5 for the $d$ = 3.8nm ribbon. As shown in figure 5e, this could be further increased to $ZT$ ~2.8 if the Fermi energy is shifted slightly by -0.03eV relative to the DFT value via doping of the graphene electrodes or by gating the device as demonstrated in [44] for single-layer MoS$_2$. This suggests that an ideal device architecture consists of multiple MoS$_2$ strips (nanoribbons) sandwiched between two graphene layers (figure S14) in which phonon transport is significantly reduced and electrons with energies close to the Fermi energy are transmitted with high probabilities through the edges of MoS$_2$ strips.

This enhancement of *ZT* for CP heterostructures is also found in other geometries. Figures S2a shows an alternative sandwich structure comprising a single layer of graphene on top of a single layer of $MoS_2$ and demonstrates that compared with parent materials, *ZT* is also increased for this structure (figure S3 and S4). Similarly figure S2b shows a variant of the structure of figure 1, whose thermoelectric properties are almost identical. This persistence suggests that enhancement of thermoelectric performance is a generic feature of such CP heterostructures. We note that although the phonon thermal conductance increases slightly with the size of the $MoS_2$ ribbons (figure S7 of the SI), the efficiency of the device is preserved since the thermal conductance is dominated by electrons in CP Gr/$MoS_2$/Gr heterostructure (figures 5b) where the phonon transmission is significantly supressed by the Gr/$MoS_2$ interface in CP heterostructure (figure S9 of the SI).

To obtain an order-of-magnitude estimate of the ribbon width beyond which high values of *ZT* may cease, we note that as shown by Figure 4, the transport resonances due to edge states are resilient, since the band of resonances just above $E_F$ change only slightly when the ribbon width is doubled. This means that transport should be dominated by edge states for a wide range of *d*. Indeed Figure S8b shows that for both ribbons, the contribution to electron transmission at energies far from resonances, (ie due to tunnelling through the bulk $MoS_2$ part of the ribbon) is approximately 100 times smaller than the contribution from the edges. Since the tunnelling current should increase in proportion to the area of the ribbon this should be negligible for ribbons up to 100 times wider than the ribbons analyzed above.

On the other hand the thermal conductance should eventually increase linearly with *d* and so at large enough *d*, *ZT*~1/κ ~1/*d*. The main question is at what value of d does this cross over take place. For small widths increasing d from 1.9nm to 3.8nm and then to 7.6nm shows that the phonon contribution increases with *d*, as expected. However at these widths, the electronic contribution dominates and therefore the total room-temperature thermal contribution decreases when *d* increases from 1.9nm to 3.8nm. This decrease is due to the narrowing of the electron transmission resonances closest to $E_F$ when *d* increases from 1.9nm to 3.8nm. This narrowing is due to the fact that the coupling between edge states decreases with *d*. For small *d*, the bonding and anti-bonding combinations possess slightly different energies and overlap to produce a more broad transmission feature, whereas at larger *d*, the edge states decouple and their transmission resonances overlap completely. At low temperatures, this narrowing decreases both *G* and $\kappa_{el}$ as expected from the Wiedmann-Franz (WF) law, but at room temperature, when the WF law is not valid, $\kappa_{el}$ decreases, while *G* increases. These changes combine to yield an overall increase in *ZT* at room temperature. Figure S7 shows that phonon thermal conductances for ribbons of width 1.9nm, 3.8nm and 7.6nm are all lower than the electronic contribution of 1.9nm and 3.8nm ribbons. Figure S12 shows that a crude extrapolation of these result to higher values of *d* suggests that phonons become dominant at ribbon widths of approximately 20nm, which are readily accessible using modern lithographic techniques.

## 3. Conclusion

In summary, we have calculated both the electron and phonon transport properties of graphene/MoS$_2$/graphene sandwich heterostructures, in which heat and electricity flows through the MoS$_2$ layer, perpendicular to the planes of the graphene electrode(s). Due to the enhanced electron transport associated the edges of ribbons, combined with phonon boundary scattering and phonon filtering at the MoS$_2$/graphene interfaces, we find that cross-plane *ZT* of the heterostructures is significantly higher than the in-plane *ZT* values of the component layers, achieving a value as high as 2.8, depending on the location of the Fermi energy and ribbon width. This demonstrates that CP heterostructures are attractive targets for the efficient conversion of waste heat into electricity via the Seebeck effect or for efficient cooling via the Peltier effect.

**Computational methods**

***Geometry optimization:*** The geometry of each structure, (graphene, MoS$_2$ and Gr/MoS$_2$/Gr heterostructure) was relaxed to the force tolerance of 10 meV/Å using the *SIESTA* [41] implementation of density functional theory (DFT), with a double-ζ polarized basis set (DZP) and the Generalized Gradient Approximation (GGA) functional with Perdew-Burke-Ernzerhof (PBE) parameterization. To check the effect of vdW interactions (see SI), van der Waals functionals (VDW) with DRSLL parameterization was used when needed. A real-space grid was defined with an equivalent energy cut-off of 300 Ry.

***Phonons transport:*** Following the method described in [36] a set of *xyz* coordinates were generated by displacing each atom from the relaxed *xyz* geometry in the positive and negative *x*, *y* and *z* directions with $\delta q' = 0.01$Å. The forces $F_i^q = (F_i^x, F_i^y, F_i^z)$ in three directions $q_i = (x_i, y_i, z_i)$ on each atom were then calculated and used to construct the dynamical matrix $D_{ij} = K_{ij}^{qq'}/M_{ij}$ where the mass matrix $M = \sqrt{M_i M_j}$ and $K_{ij}^{qq'} = [F_i^q(\delta q_j') - F_j^q(-\delta q_j')]/2\delta q_j'$ for $i \neq j$ obtained from finite differences. To satisfy momentum conservation, the *K* for $i = j$ (diagonal terms) is calculated from $K_{ii} = -\sum_{i \neq j} K_{ij}$. The phonon transmission $T_{ph}(\omega)$ then can be calculated from the relation $T_{ph}(\omega) = Trace(\Gamma_L^{ph}(\omega) G_{ph}^R(\omega) \Gamma_R^{ph}(\omega) G_{ph}^{R\dagger}(\omega))$ where $\Gamma_{L,R}^{ph}(\omega) = i(\Sigma_{L,R}^{ph}(\omega) - \Sigma_{L,R}^{ph}{}^\dagger(\omega))$ describes the level broadening due to the coupling to the left *L* and right *R* electrodes, $\Sigma_{L,R}^{ph}(\omega)$ are the retarded self-frequencies associated with this coupling and $G_{ph}^R = (\omega^2 I - D - \Sigma_L^{ph} - \Sigma_R^{ph})^{-1}$ is the retarded Green's function, where *D* and *I* are the dynamical and the unit matrices, respectively. The phonon thermal conductance $\kappa_{ph}$ at temperature *T* is then calculated from $\kappa_{ph}(T) = \frac{1}{2\pi} \int_0^\infty \hbar\omega T_{ph}(\omega) \frac{\partial f_{BE}(\omega,T)}{\partial T} d\omega$ where $f_{BE}(\omega, T) = (e^{\hbar\omega/k_B T} - 1)^{-1}$ is Bose–Einstein distribution function and $\hbar$ is reduced Planck's constant and $k_B = 8.6 \times 10^{-5}$ $eV/K$ is Boltzmann's constant.

***Electrons transport:*** To calculate electronic properties of the device, from the converged DFT calculation, the underlying mean-field Hamiltonian *H* was combined with our quantum transport code, *GOLLUM* [36,42]. This yields the transmission coefficient $T_{el}(E)$ for electrons of energy *E* (passing from the source to the drain) via the relation $T_{el}(E) = Tr(\Gamma_L^{el}(E) G_{el}^R(E) \Gamma_R^{el}(E) G_{el}^{R\dagger}(E))$ where $\Gamma_{L,R}^{el}(E) = i(\Sigma_{L,R}^{el}(E) - \Sigma_{L,R}^{el}{}^\dagger(E))$ describes the level broadening due to the coupling between left *L* and right *R* electrodes and the central scattering region, $\Gamma_{L,R}^{el}(E)$ are the retarded self-energies associated with this coupling and $G_{el}^R = (ES - H - \Sigma_L^{el} - \Sigma_R^{el})^{-1}$ is the retarded Green's function, where *H* is the Hamiltonian and *S* is the overlap matrix obtained from *SIESTA* implementation of DFT.

***Thermoelectric properties:*** Using the approach explained in [36,37,42,43], the electrical conductance $G_{el}(T) = G_0 L_0$, the electronic contribution of the thermal conductance $\kappa_{el}(T) = (L_0 L_2 - L_1^2)/hTL_0$ and the thermopower $S(T) = -L_1/eTL_0$ are calculated from the electron transmission coefficient $T_{el}(E)$ where the momentums

$L_n(T) = \int_{-\infty}^{+\infty} dE\, (E - E_F)^n\, T_{el}(E) \left(-\frac{\partial f_{FD}(E,T)}{\partial E}\right)$ and $f_{FD}(E,T)$ is the Fermi-Dirac probability distribution function $f_{FD}(E,T) = (e^{(E-E_F)/k_BT} + 1)^{-1}$, $T$ is the temperature, $E_F$ is the Fermi energy, $G_0 = 2e^2/h$ is the conductance quantum, $e$ is electron charge and $h$ is the Planck's constant. The full thermoelectric figure of merit $ZT$ is then calculated as $ZT(E_F,T) = G(E_F,T)S(E_F,T)^2 T/\kappa(E_F,T)$ where $G(E_F,T)$ is the electrical conductance, $S(E_F,T)$ is thermopower, $\kappa(E_F,T)$ is the thermal conductance due to the electrons and phonons $\kappa(E_F,T) = \kappa_{el}(E_F,T) + \kappa_{ph}(T)$, $E_F$ is Fermi energy and $T$ is temperature.

## Acknowledgment

H.S. thanks UK EPSRC for a post doctorate position funded by Quantum Effects in Electronic Nanodevices "QuEEN" platform grant no. EP/N017188/1. S.S. thanks European Commission (EC) for a Marie Curie Early Stage Researcher position within EC FP7 ITN Molecular-Scale Electronics "MOLESCO" project no. 606728. This work was also supported by UK EPSRC grant no. EP/M014452/1.

*Supporting information*

# Cross-plane enhanced thermoelectricity and phonon suppression in graphene/MoS$_2$ van der Waals heterostructures

Hatef Sadeghi[*], Sara Sangtarash and Colin J. Lambert[*]

Quantum Technology Centre, Physics Department, Lancaster University, Lancaster LA1 4YB, UK

[*]*h.sadeghi@lancastter.ac.uk; c.lambert@lancaster.ac.uk*

## Contents

## 1- Atomic structure of graphene/MoS2/graphene heterostructures

Fig S1 shows graphene-MoS$_2$-graphene heterostructure where a MoS$_2$ strip is sandwiched between two graphene layers. Graphene layers are periodic in y and z directions whereas MoS$_2$ is periodic only in z direction.

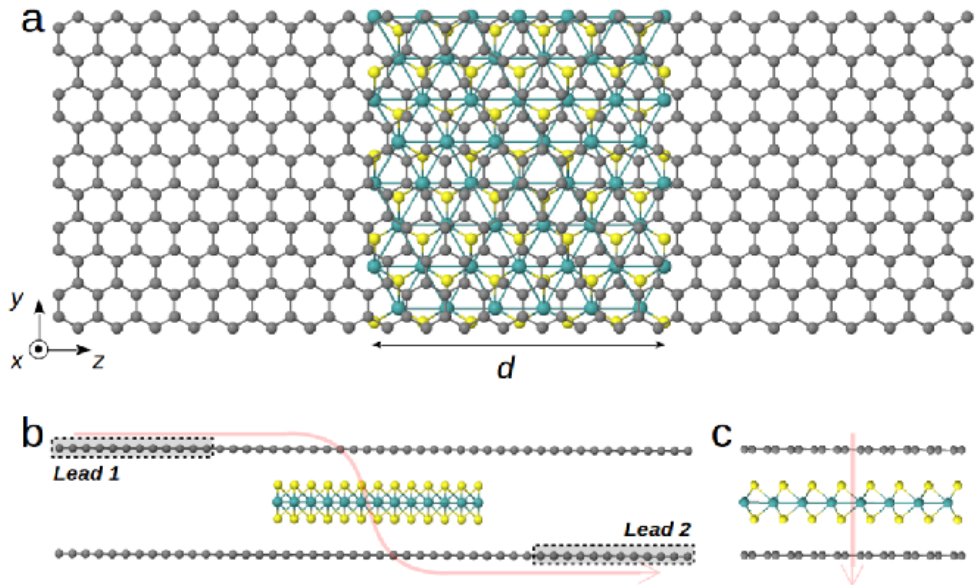

**Fig. S1. Top, side and front views of the graphene/MoS$_2$/graphene heterostructure analysed in the main text.** The red arrows shows the CPP electron and phonon transport direction.

## 2- Atomic structure of two alternative CP graphene/MoS$_2$ heterostructures

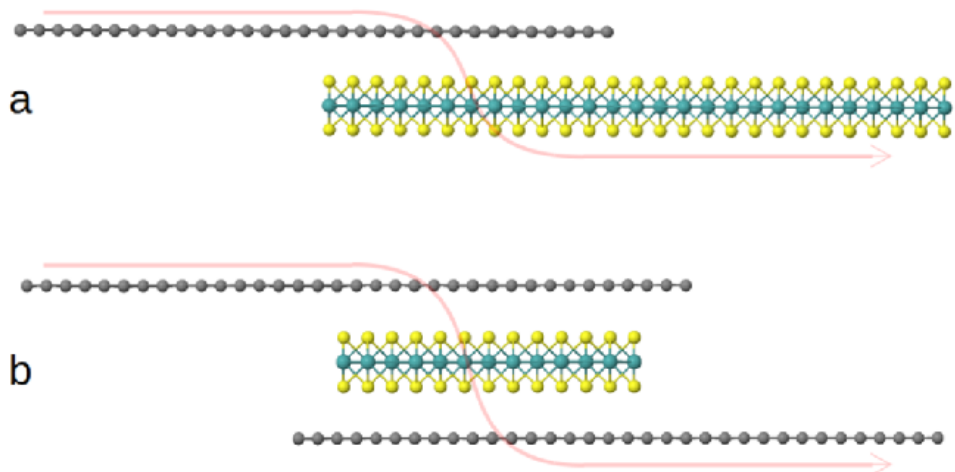

**Fig. S2.** (a) A CPP graphene-MoS$_2$ heterostructure comprising a single layer of graphene overlapping a single layer of MoS$_2$ (b) An alternative graphene/MoS$_2$/graphene heterostructure. The red arrows shows the CPP electron and phonon transport direction.

The Seebeck coefficient of structure b is almost indistinguishable from those of figure 1 of the main text.

## 3- Electron and phonon transport in graphene-MoS$_2$ heterostructures

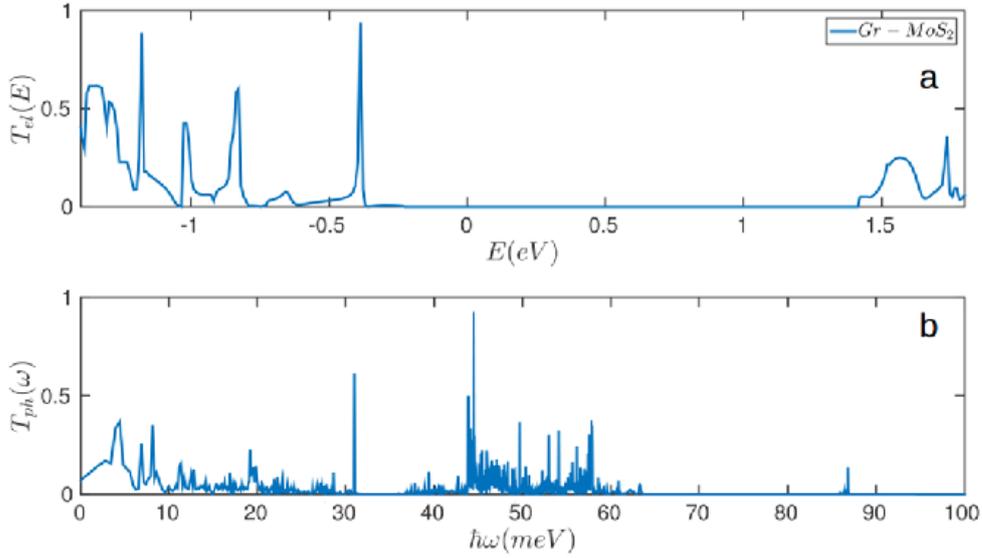

**Fig. S3. Electron and phonon transport in graphene-MoS$_2$ heterostructure of fig S2a** (a) the electron transmission coefficient $T_{el}(E)$ and (b) the phonon transmission coefficient $T_{ph}(\hbar\omega)$.

## 4- Thermal properties of graphene-MoS$_2$ heterostructures

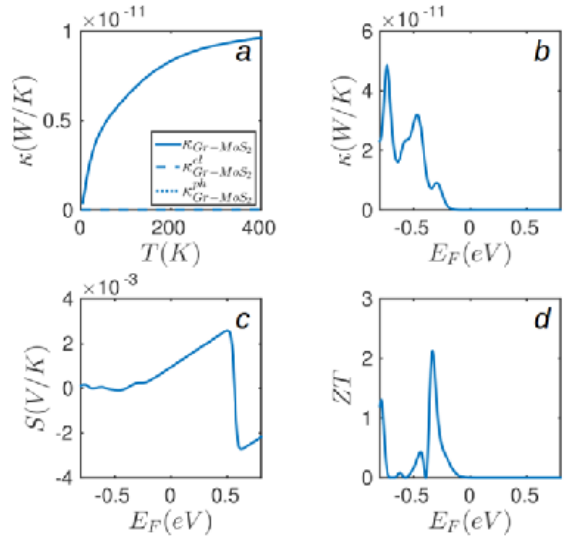

**Fig. S4. Thermoelectric properties of Graphene-MoS$_2$ heterostructure of fig S2a.** (a) The total thermal conductance $\kappa$ (solid line) and its electron $\kappa_{el}$ (dashed line) and phonon $\kappa_{ph}$ (dotted line) contribution. The room temperature Fermi energy dependence of (b) electronic thermal conductance, (c) the thermopower S and (d) total thermoelectric figure of merit $ZT$ in graphene-MoS$_2$ heterostracture. In (a) $\kappa_{el}$ is obtained at DFT given Fermi energy.

## 5- Thermoelectric properties associated with model delta-like and step-like electron transmission coefficients

To demonstrate that delta-function-like and step-function-like electron transmission coefficients lead to enhanced thermoelectric properties, the following model calculations shows how the electrical conductance, Seebeck coefficient and electronic thermal conductance depend on the width and position of such features.

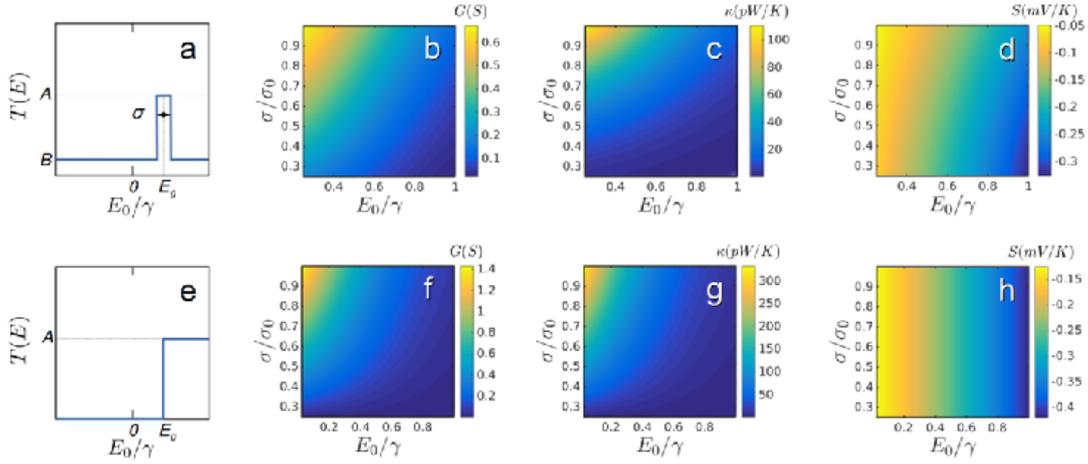

**Fig. S5. The model calculation using delta function and step function transmission coefficients $T(E)$.** (a) delta-function-like $T(E)$, (b,c,d) For the delta function like $T(E)$ of (a), the computed room-temperature electrical conductance (b), electronic thermal conductance (c) and thermopower S (d). (e) A step function like $T(E)$. For the step-like $T(E)$ of (e), the computed room-temperature electrical conductance (f), electronic thermal conductance (g) and thermopower S (h). Note that $\gamma = \sigma_0 = 4k_BT$.

## 6- Local density of states in the vicinity of Fermi energy in the graphene/MoS$_2$/graphene heterostructure of figure S1

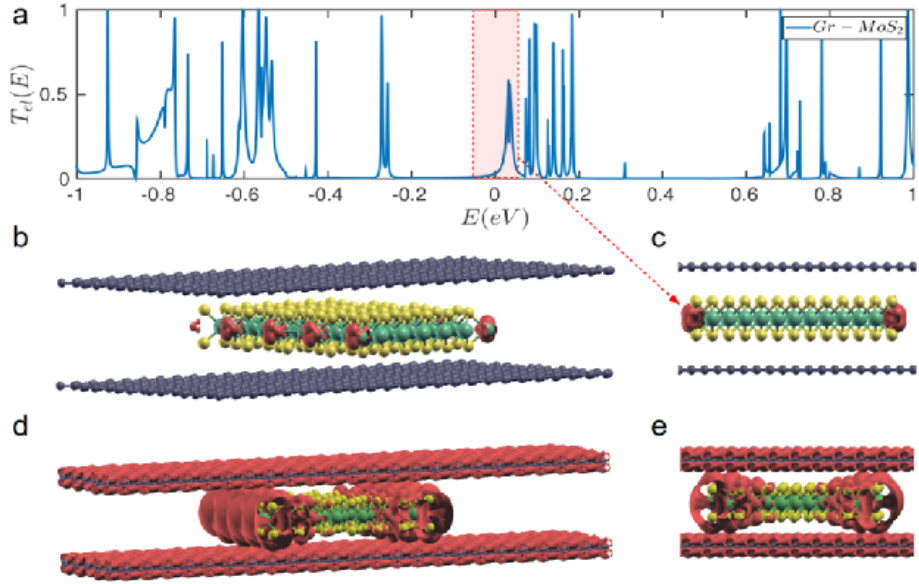

**Fig. S6. Local density of states in the vicinity of Fermi energy $E_F = 0$ eV in graphene/MoS$_2$/graphene heterostructure of fig S1, with d=1.9nm.** (a) the electron transmission coefficient $T_{el}(E)$. (b,c) local density of states from two different view in the vicinity of Fermi energy shown by red area in (a) with small iso-surface. (d,e) local density of states from two different views in the vicinity of Fermi energy shown by red area in (a) with larger iso-surface.

## 7- Effect of increasing the width of the MoS$_2$ ribbons (*d* in figure S1)

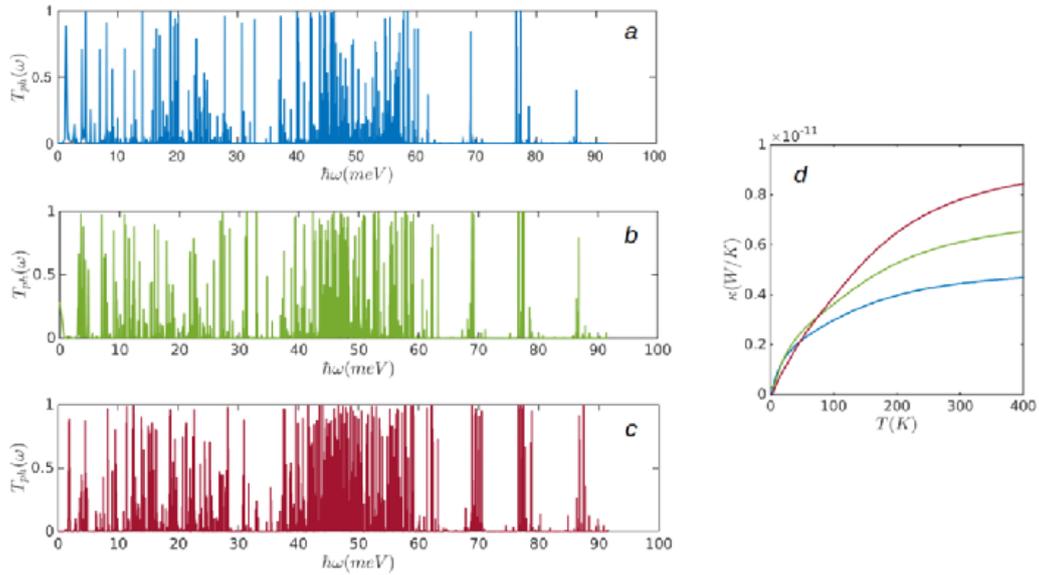

**Fig. S7. Phonon transmission through the graphene/MoS$_2$/graphene structure with two different sizes of MoS$_2$ strips.** (a) for the structure shown in fig. S1 of width 1.9nm, (b) for the structure shown in fig. S1 where the MoS$_2$ is of width 3.8nm, (c) for the structure shown in fig. S1 where the MoS$_2$ is of width 7.6nm and (d) phonon thermal conductance in these structures.

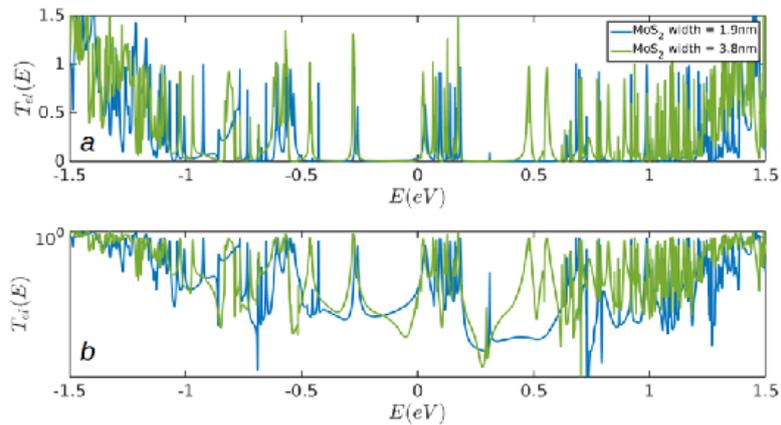

**Fig S8. The electron transmission coefficients.** For ribbons with d= 1.9nm and d=3.8nm plotted over a wide energy range, showing the locations of the MoS$_2$ valence and conduction band edges around E= -1eV and E=1eV (a) in normal scale and (b) in logarithmic scale.

## 8- Phonon transmission coefficient in Gr/MoS$_2$/Gr compared with pristine graphene and MoS$_2$

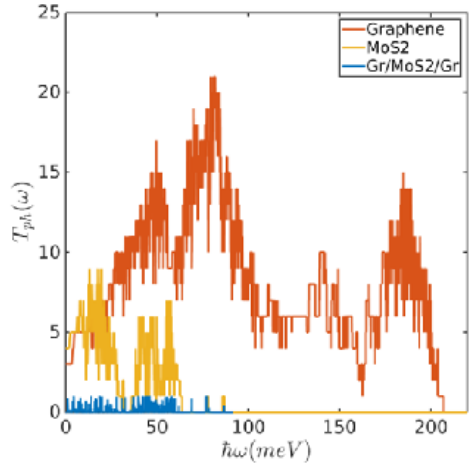

**Fig. S9.** Phonon transport in in-plane pristine graphene and MoS$_2$ and in the cross plane graphene/MoS$_2$/graphene heterostructure.

## 9- Van der Waals interactions

DFT calculations using van der Waals functional is computationally expensive and specifically in the systems discussed in this paper with more than 1000 atoms, it makes the calculation not feasible. This is even more expensive for phonon calculations since for three systems with each about 1000 atoms in which 6 calculations per atom needed, totally 3×6000=18000 independent calculations should be carried out. Each calculation takes approximately on average 6 hours on a 16 processor node. In total 1,728,000 CPU hours is needed just for phonon thermal conductance calculations. However as discussed below, we could take the effect of van der Waals interaction into account by appropriate approximations

As a first step, we computed the cross-plane electronic properties using a vdW functional [1,2] using the MoS$_2$/graphene ground state geometry obtained from a GGA-PBE. As shown in the figure S10 (blue and red graphs) the transmission coefficient changed only slightly. The main effect is increment of the energy gap by 0.5 eV. However, in the most relevant energy intervals [0, 0.2eV] around DFT-predicted Fermi energy, the transmission is not changed significantly. This shows that for a given geometry, the GGA-PBE and vdW functionals yield similar mean-field Hamiltonians. However as shown in the table S1, the interlayer distance between graphene and MoS$_2$ is slightly larger with vdW functionals.

**Table S1: MoS$_2$ – graphene interlayer distance**

| Ref | LDA | GGA | vdW | code |
| --- | --- | --- | --- | --- |
| Ghorbani-Asl, *et al.* [3] | - | 3.192 | - | Siesta |
| Jin, *et al.* [4] | 3.3 | - | 3.4 | GPAW |
| Le, *et al.* [5] | - | 4.32 | 3.54 | VASP |
| Liu, *et al.* [6] | - | - | 3.37 | VASP |
| Ma, *et al.* [7] | 3.32 | - | - | VASP |
| Shao, *et al.* [8] | 3.32 | - | - | VASP |
| Wang, *et al.* [9] | 3.32 | - | 3.32 | VASP |
| *Our work* | - | 3.4 | - | Siesta |

To study the effect of this increase, we increased the interlayer separation between $MoS_2$ and graphene by 0.3 Å and calculated $T(E)$ using the GGA-PBE functional (orange graph in the figure S10). By increasing the interlayer distance, the overlap between the pi orbitals of graphene and $MoS_2$ decreases and therefore, the conductance decreases. This simultaneously decreases the electronic conductance and thermal conductance due to the electrons and improves the thermopower, since the slope of the logarithm of the transmission function *ln T(E)* close to the DFT Fermi energy ($E_F = 0$) increases.

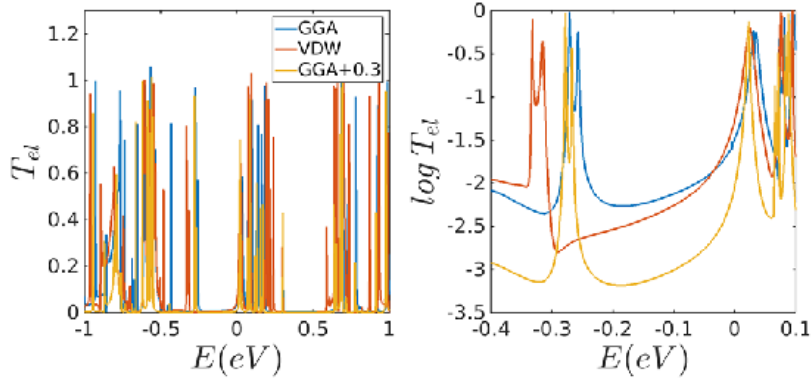

**Fig. S10. Electron transmission coefficient through the cross plane graphene/MoS2/graphene heterostructure.** A comparison between $T_{el}(E)$ using GGA-PBE functional in this paper, vdW functional with the optimized geometry obtained by GGA-PBE functional and GGA-PBE functional with increased (by 0.3Å) interlayer distance between graphene/$MoS_2$.

Figure S11 shows the thermal properties of the junction using *T(E)* of figure S10a. The changes in the transmission coefficient have some effect on the thermal properties. Around the DFT predicted Fermi energy $E_F = 0$, *ZT* is enhanced mainly due to an increase in the thermopower. This demonstrates that if vdW funtionals are used the predicted efficiency of the device improves. Therefore, our calculation in this paper may be an underestimate of *ZT*.

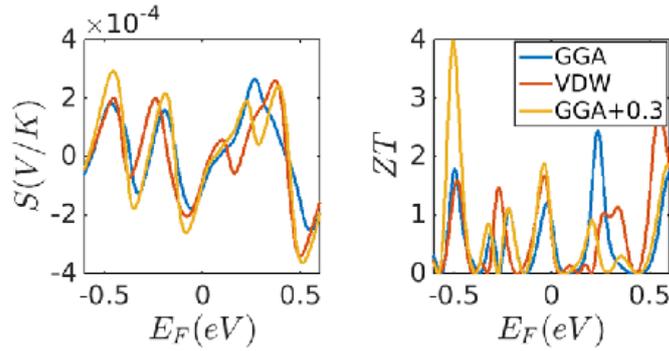

**Fig. S11. Thermoelectric properties of cross plane graphene/MoS2/graphene heterostructure.** Thermopower (left) and thermoelectric figure of merit ZT (right) for the transmission coefficients shown in fig S10 to account for using different functionals.

## 10- How does ZT enhancement scale with the MoS$_2$ ribbon width?

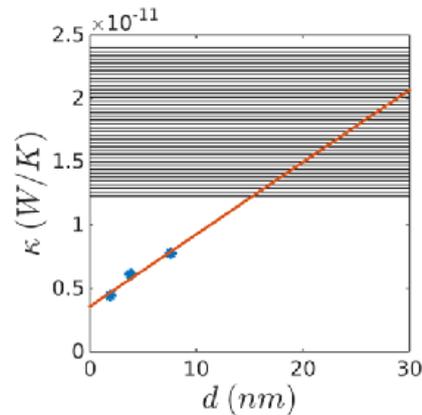

**Fig. S12. Thermal conductance in the graphene/MoS2/graphene structure with three different sizes of MoS$_2$ strips.** Phonon thermal conductance for ribbons with width of $W$ = 1.9nm, 3.8nm, 7.6nm (blue dots) and its straight line fit (red line). The upper and lower bounds of the shaded regions show electron thermal conductance around the Fermi energy at room temperature for ribbons of width 1.9nm and 3.8nm. Phonons begin to dominate for ribbons wider than approximately 20nm.

## 11- Electrical conductance G/G$_0$ from *T(E)*

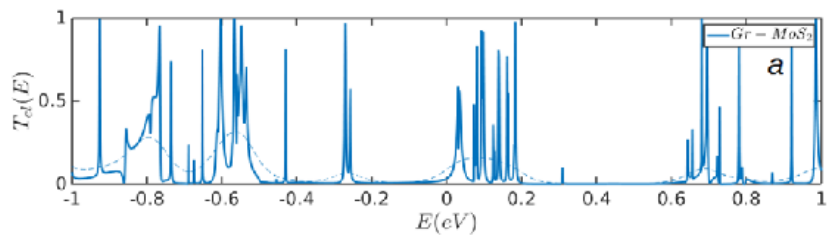

**Fig S13. As an example of the relationship between electrical conductance and transmission**, the solid lines in this figure show the electron transmission coefficient of the structure containing a 1.9nm ribbon, while the dashed line shows the corresponding room-temperature electrical conductance in units of the conductance quantum.

## 12- Cross-plane thermoelectric device architecture

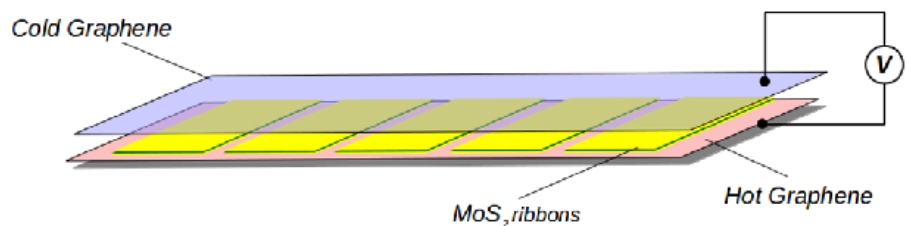

**Fig S14. An example of cross-plane thermoelectric device architecture** using van der Waals graphene/MoS2/graphene heterostructures consisting of multiple MoS$_2$ strips (nanoribbons) sandwiched between two graphene layers.

## 13- References